\documentclass[12pt,a4paper]{article}
\usepackage[T2A]{fontenc}
\usepackage[cp1251]{inputenc}
\usepackage[russian,english]{babel}
\usepackage{amssymb}
\usepackage{amsmath}
\usepackage{graphicx}
\usepackage[small,sc,center]{titlesec}
\usepackage{indentfirst}
\usepackage[labelsep=period]{caption}
\usepackage{caption}
\newcommand{\msun}{\,$M_{\odot}$}
\newcommand{\msyr}{\,$M_{\odot}$\,yr$^{-1}$}

\newcommand{\ergcms}{\,erg\,cm$^{-3}$\,s$^{-1}$}
\newcommand{\ergs}{\,erg\,s$^{-1}$}

\newcommand{\kms}{\,km\,s$^{-1}$}

\newcommand{\cmsqg}{\,cm$^2$\,g$^{-1}$}

\newcommand{\ha}{H$\alpha$}

\newcommand{\feii}{Fe\,{\sc ii}}
\newcommand{\oi}{O\,{\sc i}}

\newcommand{\caii}{Ca\,{\sc ii}}
\newcommand{\cii}{C\,{\sc ii}}

\begin{document}
	
\begin{center}
\textbf{SN~Ia-CSM 2020aeuh: Massive binary C/O WD merger?}

\vskip 5mm
\copyright\quad
2025 \quad N. N. Chugai \footnote{email: nchugai@inasan.ru}\\
\textit{Institute of astronomy, Russian Academy of sciences, Moscow} \\
Received  10.09.2025
\end{center}

{\em keywords:\/} stars  -- supernovae; stars -- binary stars 

\noindent
{\em PACS codes:\/} 

\clearpage
 
 \begin{abstract}
\noindent
I explore the origin of the circumstellar (CS) shell of the unusual SN~Ia 2020aeuh 
 based on the light curve model abd observational constraints. 
I estimate the $^{56}$Ni mass (1.1\msun), CS shell mass (0.04--0.2\msun), radius 
($2\times10^{16}$\,cm), and expansion velocity $\lesssim 200$\kms.
Large $^{56}$Ni mass and properties of the CS shell are consistent with the scenario 
  of massive binary white dwarf merger that has been accompanied with the ejection of 
   $\sim 0.1$\msun\ of matter.  
It is argued that SN~2020aeuh exploded not earlier than 30 yr after the merger.   
\end{abstract}

\section{Introduction}

A variety of supernovae SN~Ia interacting with a dense circumstellar matter (CSM)  came 
   into limelight after the detection of SN~2002ic with  
   the SN~1991T-like spectrum  (Hamuy et al. 2003) and the circumstellar (CS) \ha\ emission  with the 
   apparent signature of the CS interaction. 
Currently, the category of SN~Ia-CSM comprises 28 events (Sharma et al. 2023).
In all the cases spectra show CS emission lines of hydrogen and/or helium; this fact 
 presumably indicates a single degenerate scenario for these SN~Ia.

  Unusual SN~2020aeuh studied in detail by Tsalapatas et al (2025) belongs to this subclass, but it stands out by two spectacular features:
  (i) the CS interaction turned on only at about +50 days after the discovery, so the initial luminosity maximum of SN~Ia  on +15 day is well seen being separated from the CS interaction luminosity bump; (ii) the CSM does not show either hydrogen or helium  emission lines; instead it demonstrates  \oi\ 7774\,\AA\ and \cii\ 5890\,AA\ narrow emission lines impying C/O-rich matter (Tsalapatas et al. 2025). 
Authors estimate the mass of the CS shell to be 1-2\msun\  and the radius of $\approx10^{16}$\,cm (Tsalapatas et al. 2025).
 
The \oi\ and \cii\ emission lines from CSM compells ones to propose the origin of the CSM in the scenario of double C/O white dwarf (WD) merger (Tsalapatas et al. 2025, Soker 2025)
 \footnote{SN~I via double C/O WD merger (cf. Webbink 1979)}. 
Yet authors acknowledge (Tsalapatas et al. 2025) that
 large CSM mass is difficult to reconcile with this scenario. 
A 3D-simulation of binary WDs fly-by collision in parabolic orbits results in 
 the ejection, at best, of a few tenths of solar mass, far below required amount ($>1$\msun) (Tsalapatas et al. 2025). 
  
On the other hand, the high maximum luminosity on day +15 (-19.7 mag) means that SN~2020aeuh 
belongs to a family of overluminous SNe~Ia that suggests a merger of massive C/O WDs 
(Taubenberger et al. 2013).
The ejection of a few tenths of solar mass during the merger of massive C/O WD, e.g.,  with masses
1\msun\ + 0.8\msun,  might be an appropriate scenario for SN~2020aeuh provided the 
  actual CS mass significantly smaller than the solar mass.
  
The challenging problem of the  origin of the CSM around  SN~2020aeuh emphasises the significance of 
   the CSM mass estimate as a possible telltail of the underlying mass-loss mechanism.
The  goal of this paper is an alternative estimate of the  
CS shell mass based of the  CS interaction modelling. 
It turns out that the CSM mass is a small fraction of the solar mass, so   
 the double C/O WD merger scenario for SN~2020aeuh seems plausible.

\section{Preliminaries}

Spectra obtained at the CS interaction stage ($\geq +81$ day) are similar to the spectrum of SN~2002ic (disregarding \ha) on day 244 (Wang et al. 2004): 
 they are composed by broad emission bands,  primarily of \feii\ lines, that originate from a fragmented  cold dense shell (CDS) formed between forward and reverse shocks (Chugai et al. 2004). 
The \caii\ emission lines both ultraviolet doublet and infrared triplet  
permit ones to infer the expansion velocity of the CDS. 
Clear-cut blue edge of \caii\ 3933\,\AA\ on day +99 (cf. Tsalapatas et al. 2025) suggests the CDS expansion velocity of $\approx 11000$\kms.
This important obeservable is used below to additionaly constrain the CS interaction model. 

Less obvious  is the issue of the CSM expansion velocity. 
The \oi\ 7774\,\AA\ emission with the FWHM of $\sim 1300$\kms\ (Tsalapatas et al. 2025)  is apparently related to the CS shell. 
The narrow core  ($\sim200$\kms) and broad wings ($\pm2000$\kms) of \oi\ 7774\,\AA\ 
 remind CS emission lines from a slowly expanding shell  
 with a large Thomson optical depth ($\tau_{\tiny{T}} > 1$), similar to early SN~1998S 
  (Chugai 2001).
However, for SN~2020aeuh this broadening  mechanism is rulled out, because for 
  the shell with the radius of $10^{16}$ cm and 1\msun\ mass composed of carbon and oxygen 
  the Thomson optical depth is small, $\tau_{\tiny{T}} < 0.1$ (for the double ionization of C and O).  
  
%================================================================
\begin{figure}
	\centering
	\includegraphics[trim=0 100 0 100, width=\columnwidth]{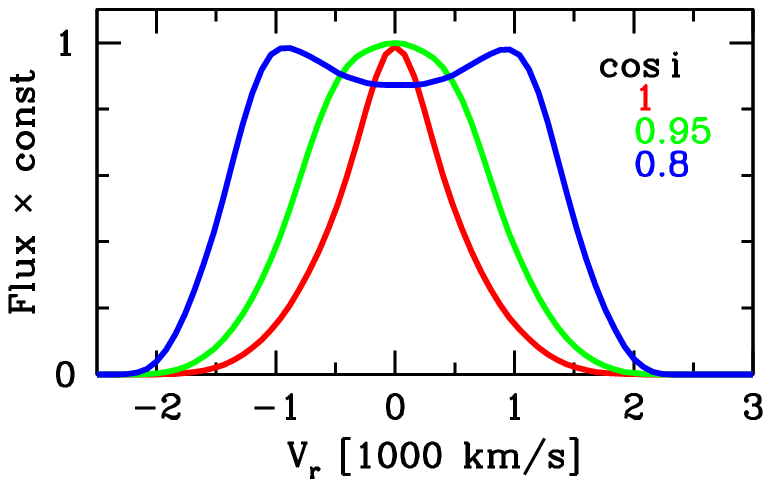}
	\caption{
		Theoretical profile of a line emitted by the CS shell with the spherical velocity 
		distribution, but non-spherical density distribution of equatorial type 
		with the emissivity depending on the polar angle $\theta$ as $\epsilon \propto (1 - |\cos{\theta}|)^3$ for different 
		cosine of an inclination angle. 
		} 
	
	\label{fig:int}
\end{figure}
%======================================================   

Another possibility is that the  \oi\ 7774\,\AA\ line forms due to the supernova interaction 
 with a slowly expanding clumpy CSM likewise, e.g.,
  the intermediate width \ha\ of IIn supernova SN~1988Z (Chugai \& Danziger 1994).
  The crushed CS clouds fragmented and accelerated in the forward shock (Klein et al. 1994)   
  produce a broad velocity spectrum of the line-emitting gas, which also has been suggested for the SN~2002ic \ha\ profile (Chugai \& Chevalier 2007).
In the framework of this mechanism the preshock expansion velocity of the clumpy shell 
   in SN~2020aeuh is presumably $\lesssim200$\kms\ (the upper limit is the spectral resulution).
   
Remarkably, admitting CSM anisotropy one can produce the line profile similar to \oi\  
  7774\,\AA, if one assumes a spherical expansion of the CSM with a constant velocity of 2000\kms\ and axially symmetric matter distribution with the density maximum at the equator.
To illustrate the point we adopt the line emissivity (\ergcms) distribution aong the polar 
   angle $\theta $ to be $\epsilon  \propto (1 - |\cos{\theta}|)^q$. 
For $q = 3$ the line profile observed from the polar direction with zero inclination 
  ($\cos{i} = 1$) (Fig. 1) is indeed similar to the observed \oi\ 77774\,\AA\ line on day +99 
  (cf. Tsalapatas et al. 2025).
However, similarity disappears even for a small deviation from the polar zxis.
For $\cos{i} = 0.95$ the profile loses sharp vertex, while for the larger deviation from the polar 
 axis ($\cos{i} = 0.8$) the profile acquires two horn shape  -- classic signature of 
 equatorial belt.
Small probability of the favorable line of sight ($< 0.05$)  makes this anisotropic model 
 interesting, but improbable for SN~2020aeuh.
A spherical clumpy CS shell thus is a more appropriate assumption for SN~2020eauh.
 
The  total radiated energy of the second bump of SN~2020aeuh powered by the CS interaction  is $Q = 1.1\times10^{50}$\,erg  (Tsalapatas et al. 2025).
This energy is released in the forward and reverse shock as a thermal energy, 
   followed by the X-ray radiation of the hot gas; X-rays are partially absorbed in the unshocked 
    ejecta, CDS, and CSM with the subsequent instant conversion into the observed optical radiation.
    (the diffusion time is small).
    
An additional mechanism of the shock wave energy conversion into the optical radiation is related to the CDS Rayleigh-Taylor instability that brings about the CDS fragmentation
   and cold matter mixing with the hot gas of the forward shock (cf. Blondin and Ellison 2001). 
The thermal conductivity heats up the cold CDS fragments and thus additionally  converts the 
   shock wave energy into the optical radiation.
   
The lower limit of the CSM mass can be estimated assuming the inelastic collision\footnote{i.e., total conversion of kinetic energy into radiation in a zero momentum frame} of 
  ejecta external mass $m_1$ with the CSM of mass $m_2$.
The momentum and energy conservation results in the radiated energy 
  $Q = 0.5m_2u^2(1 + m_2/m_1)$, where $u = 11000$\kms\ is the expansion velocity of the CDS with the mass $m_1 + m_2$.
 For the reasonable assumption $m_1 \approx m_2$ (supported by the CS interaction model) the above expression  with the known $Q$ results in the  CSM mass lower limit $M_{cs} >  m_2 \approx 0.04$\msun.

 % ==================================================================
\vspace{1cm}
\begin{table}[hb]
	\centering
	\caption{Model parameters}
	\begin{tabular}{ccccc} 
		\hline
		 $M$  & $E$ &  $M_{ni}$   & $M_{csm}$  & $R_{cs}$     \\
		 (\msun)  & ($10^{51}$\,erg) &  (\msun) & (\msun) & ($10^{16}$\,cm)   \\

		\hline 
		
		   1.4          &     1.3  &  1.1   &  0.18 & 1.85\\
		   	
			\hline
	\end{tabular}
\end{table}
%=========================================================================== 

\section{Mass of CS matter} 

The SN/CSM interaction is modelled in the thin shell approximation (Chugai 2001). 
The X-ray luminosity of the forward and reverse shock is calculated in terms of the radiation 
   efficiency of the postshock gas $\eta = t/(t + t_c)$, where $t$ is the age, and $t_c$ is the cooling time for the cooling function of Sutherland \& Dopita (1993) and assuming 
   the full equilibration of electron and ion temperatures.
 X-ray radiation of both shocks is partially absorbed by unshocked ejecta, cold dense shell (CDS), and unshocked CSM; the absorbed X-ray power is instantly converted into the observed 
  bolometric luminosity.  
The model light curve includes the luminosity of the SN~Ia powered by the radioactive decay and calculated in the Arnett approximation (Arnett 1980) assuming the opacity of 0.1\cmsqg. 
The  SN~Ia density is approximated by the exponent (Dwarkadas and Chevalier 1998) $\rho = \rho_0\exp{(-v/v_0)}$ 
  where $\rho_0$ and $v_0$ are defined by the ejecta mass $M$ and kinetic energy $E$ 
  taking into account homologous expansion, $\rho_0 \propto 1/t^3$.
For the CSM of the mass $M_{cs}$ the adopted radial density distribution is a  
  broken power law with the constant expansion velocity of 200\kms.

The shown results (Fig. 2) suggest parameters of Table 1 that
 includes the adopted ejecta mass and energy, as well as the inferred  $^{56}$Ni mass, the CSM mass, and the CS shell radius, i.e., the density maximum (see inset).
 The uncertainty of inferred values  does not exceed 10\%.
 The model satisfactorily fits to the observed bolometric light curve and the model CDS expansion velocity
 is consistent with the CDS velocity  derived from the \caii\ line on day 99 
 (Fig. 2).

The forwatd shock, generally, is able to additionally contribute in the bolometric  
 lumunosity by means of a thermal conductivity between the hot gas and 
  admixed fragments of the CDS as well as shocked CS clouds.
  This additional luminosiry can result in the smaller CSM mass compared to the value inferred 
   by the CS interaction model. 
 The actual CSM mass, therefore, lies in the range $0.04 < M_{cs} < 0.2$\msun.
    
The large amount of $^{56}$Ni ($\approx 1$\msun)  suggests that SN~2020aeuh belongs to the variety 
of overluminous SN~Ia and therefore originates likely from a 
    merger of two massive C/O WDs (cf. Taubenberger et al. 2013).
Remarkably, that even the upper value of the CS mass of 0.2\msun\ is consistent with this scenario.   
Disregarding a specific mass-loss mechanism at/after  the WD merger, a binary, say, 
of 1\msun\ + 0.7\msun\  C/O WD might have been the SN~2020aeuh progenitor.
   
 .%================================================================
 \begin{figure}
 	\centering
 	\includegraphics[trim=0 100 0 100, width=\columnwidth]{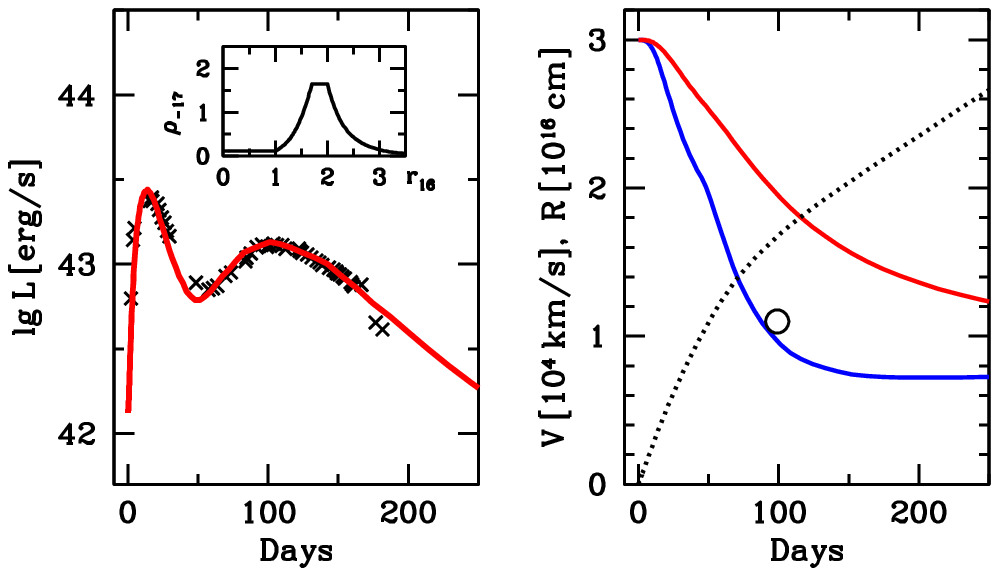}
 	\caption{
 		{\em Left.} Model bolometric light curve (red line) overplotted on observational data (Tsalapatas et al. 2025). 
 		Inset shows CSM density distrbution.	
 		{\em Right.}  The CDS velocity ({\em blue}) and the radius ({\em dotted}); maximum velocity of unshcked ejecta is shown by {\em red} line. 
 	The CDS velocity is consistent with the CDS velocity value inferred from the \caii\ 3933\,\AA\ line on day +99 (circle).	
 	}
 	\label{fig:int}
 \end{figure}
 %======================================================

\section{Discussion}

The goal of the paper has been the estimate of the CSM mass around SN~2020aeuh and  
  the anlysis of a possibility that this supernova could be the outcome of the double C/O WD merger. 
The inferred  mass of the CS shell is found to be in the range  of 0.04--0.2\msun.  
This estimate taken with the large $^{56}$Ni mass of 1\msun\ makes the
 merger of two massive C/O WD accompanied by the ejection of $\sim 0.1$\msun\ 
  a favorable scenario for SN~2020aeuh.
 The expansion velocity of the CS shell is argued to be $u \lesssim 200$\kms.
  This value along with the inferred radius of the CS shell $R_{cs} \approx 2\times10^{16}$\,cm 
  implies the explosion delay after the merger of $t_{del} = R_{cs}/u \gtrsim 30$\,yr.
  
A theory so far is not able to predict the SN~Ia explosion delay time  after the C/O WD merger 
 except for the case when the model explosion occures during the merger. 
The 3D-hydrodynamic modeling with the primary mass of 0.9\msun\ and 1.2\msun\ 
(Fenn et al. 2016) results in the explosion during the merger in the massive primary case (1.2\msun) only.  
This may be considered as an indication that for the more massive primary C/O WD the explosion delay time is shorter.   
  
If this trend reflects the reality, then  most of SNe~Ia produced by the merger of C/O WD 
    with the primary of a  moderate mass 
    explode with large delay time ($>10^2$ yr), so even, if some amount of the WD matter is ejected,  the CS interacion stage can avoid detection because of the significant expansion of the CS shell before the explosion happens.
Noteworthy, that the ejected mass could be, nevertheless,  detected via the CS shell narrow absorption lines in  SN~Ia spectra (Raskin and Kasen 2013).
 
 The loss of $\sim 0.1$\msun\ during the merger of massive C/O WD is a serious 
 problem.  
 3D-hydrodynamic simulations of the WD merger (Raskin \& Kasen 2013) have found that in all the studied cases the mass of the ejected tidal tail is $<5\times10^{-3}$\msun.
On the other hand, the detailed anlysis of the electromagnetic effects of the neutron stars (NS)
  merger in the GW170817 event indicates that the mass of ejected material is 
   $\approx 0.11$\msun\ (Vieira et al. 2025), i.e., about 10\% of the neutron star mass.
Based on the analogy between merger process in the casees of WD and NS one expects that the 
  ejection of $\sim 0.1$\msun\ during a double WD merger is plausible. 

Given the CS shell extention of $\Delta r/r \sim 0.5$ (Fig. 2), the CS shell formation in SN~2020aeut took 
 about $\sim 15$ yr with the average mass loss rate $\dot{M} \sim 7\times10^{-3}$\msyr. 
For the kinetic energy of the CS shell $\approx 4\times 10^{46}$ erg, the average 
  kinetic luminosity of the flow was therefore $\approx 10^{38}$\ergs.
Since the expansion velocity of the shell $u \sim 200$\kms\ is, presumably, of the order of the escape velocity at the outflow radius, one obtains the radius at the mass outflow  $GM/u^2 \sim 5\times 10^{12}$ cm. 

These estimates lead us to the following picture of the CS shell formation and 
the SN~2020aeuh event. 
During the violent merger a super-Chandrasekhar WD formation releases the binding energy of $\approx10^{51}$ erg. 
About 10\% of this energy is spent on the ejection of $\sim 0.1$\msun, which forms an extended atmosphere 
  with the radius of $ \sim 5\times 10^{12}$ cm.
Over the next $\sim 15$ years, the WD spin-down luminosity maintains the extended atmosphere 
 in the quasi-hydrostatic equilibrium with the mass loss rate of $\dot{M} \sim 7\times10^{-3}$\msyr.
Rougly 30 years after the merger, the SN~Ia explodes as SN~2020aeuh.

\section{Conclusion}

The paper studies the origin of the CS shell around unusual SN~Ia 2020aeuh with the 
conclusion that the merger of massive C/O WD could bring about observed phenomena.
The conclusion is based on the following results.

\begin{itemize}
	\item The mass of the CS shell lies in the range $0.04-0.2$\msun, and the shell radius 
	 is $\approx2\times10^{16}$ cm. 
	\item 
	The synthesised $^{56}$Ni amount is rather high  $\approx 1$\msun.
	\item The expansion velocity of the CS shell is $\lesssim 200$\kms, significantly lower than the  width of the \oi\ 7774\,\AA\ CS line.
	\item SN~2020aeuh exploded not earlier than 30 yr after the merger.
	
\end{itemize}

\section{Acknowledgements}

I am gratefull to Maxim Barkov for discussion and for interesting case of non-spherical 
CD shell.

%\clearpage

\section{References}

\noindent
 Arnett W. D. , Astrophys. J. {\bf 237}, 541 (1980)\\
\noindent
Chugai N. N., Chevalier R. A., Astrophys. J. {\bf 657}, 278 (2007)\\ 
\medskip
Chugai N. N., Chevalier R. A., Lundqvist P. Mon. Not. R. Astron. Soc. {\bf 355}, 627 (2004)\\
\medskip
Chugai N. N., Mon. Not. R. Astron. Soc. {\bf 326}, 1448 (2001)\\
\medskip
Chugai N. N., Danziger I. J., Mon. Not. R. Astron. Soc. {\bf 268}, 173 (1994)\\
\medskip
Dan M., Rossvoog S., Br\"{u}ggen M., Podsiadlowski Ph. Mon. Not. R. Astron. Soc. {\bf 438}, 14 (2014)\\
\medskip
Dwarkadas V.K, and Chevalier R. A., Astrophys. J. {\bf 497}, 807 (1998)\\
\medskip
Fenn D., Plewa T., Gawryszczak A. Mon. Not. R. Astron. Soc. {\bf 462}, 2486 (2016)\\
\medskip
Hamuy M., Phillips M. M., Suntzeff N. B. Nature {\bf 424}, 651 (2003)\\
\medskip
Klein R. I., McKee C. F., Colella P.  Astrophys. J. {\bf 420}, 213 (1994)\\
\medskip
Raskin C.,  Kasen D.  Astrophys. J. {\bf 772}, 1 (2013)\\ 
\medskip
Sutherland R. S., Dopita M. A.. Astrophys. J. Suppl. Ser. {\bf 88}, 253 (1993)\\
\medskip
Sharma Y., Sollerman J., Fremling C. et al.  Astrophys. J. {\bf 948}, 52 (2023)\\    
\medskip
Soker N. arXiv:2507.16757 (2025)\\
\medskip
Taubenberger S., Kromer M., Hachinger S. et al.  Mon. Not. R. Astron. Soc. {\bf 432}, 3117 (2013)\\
\medskip
Tsalapatas K., Sollerman J., Chiba R. et al. arXiv:2507.08532 (2025)\\
\medskip
Webbink R. F., {\em  White Dwarfs and Variable Degenerate Stars,  Proceedings of IAU Colloq. 53} (Ed. H. M. van Horn, V. Weidemann, University of Rochester, 1979), p.426.\\
\medskip
Vieira N.,  Ruan J. J., Haggard D. et al., arxiv: 2504.10696 (2025)\\
\medskip
Wang L., Baade D., H?flich P. et al., Astrophys. J. {\bf 604}, L53 (2004)\\
\end{document}